\documentclass[11pt]{article}
\usepackage{moriond}
\usepackage{graphicx}
\usepackage{caption}

\newcommand{\gevsq  }{\mbox{$\rm GeV^2$}}
\newcommand{\aem    }{\mbox{$\alpha$ }}
\newcommand{\qsq    }{\mbox{$Q^{2}$}}
\newcommand{\ftg    }{\mbox{$F_{2}^{\gamma}$}}
\newcommand{\ftp    }{\mbox{$F_{2}^{p}$}}
\newcommand{\ftn    }{\mbox{$F_{2}^{\gamma}/\aem$}}
\newcommand{\aemsq  }{\mbox{$\aem^2$}}

\newcommand{\qzm    }{\mbox{$\langle \qsq \rangle$}}
\newcommand{\dstar    }{\mbox{$D^{*\pm}$}}
\newcommand{\ft     }{\mbox{$F_{2}^{\gamma}$}}
\newcommand{\fl     }{\mbox{$F_{\mathrm{L}}^{\gamma}$}}
\newcommand{\flxq   }{\mbox{$\fl(x,\qsq)$}}
\newcommand{\ftxq   }{\mbox{$\ft(x,\qsq)$}}
\def\CF2{F_{\rm 2,c}^{\gamma}}
\def\DST{{\rm D}^{\ast}}
\def\ptdst{p_{\rm T}^{\DST}}

\def\XQCF2{F_{\rm 2,c}^{\gamma}(x,\langle Q^2\rangle)}
\newcommand{\epem}{$\mbox{e}^+\mbox{e}^-$}
\def\ccbar{\mbox{c}\overline{\mbox{c}}}
\def\bbbar{\mbox{b}\overline{\mbox{b}}}
\def\Journal#1#2#3#4{{#1} {\bf #2}, #3 (#4)}

\def\NPB{{\em Nucl. Phys.} B}
\def\PLB{{\em Phys. Lett.}  B}

\def\PRD{{\em Phys. Rev.} D}

\def\EPJ{{\em Eur. Phys. J.}}

\def\be{\begin{equation}}
\def\ee{\end{equation}}
\def\bea{\begin{eqnarray}}
\def\eea{\end{eqnarray}}

\begin{document}
\begin{flushright}
{\large OPAL CR-430}
\end{flushright}
\vspace*{4cm}
\title{THE PARTON CONTENT OF THE PHOTON}

\author{ THORSTEN WENGLER }

\address{CERN/EP \\CH-1211 Gen\`eve 23, Switzerland\\
         E-mail: Thorsten.Wengler@cern.ch}

\maketitle\abstracts{
The structure of the photon is studied in
photon-photon collisions at LEP. New measurements have become
available exploring the structure function \ftg\ of the photon 
to the lowest values of $x$ yet, where the hadronic component of the
photon is most important. Inclusive cross sections for \dstar-mesons have
been measured  as a function of their transverse momenta and are used
to test the validity of perturbative QCD in NLO in this region. The
total charm quark production cross section is extracted. For the first
time the inclusive production of \dstar mesons in deep inelastic 
electron-photon scattering has been used to extract the charm 
structure function of the photon. A first sign of beauty production in
photon-photon collisions is reported.
}

\section{Introduction}

\begin{figure}[ht]
\begin{center}
\includegraphics[width=0.73\textwidth]{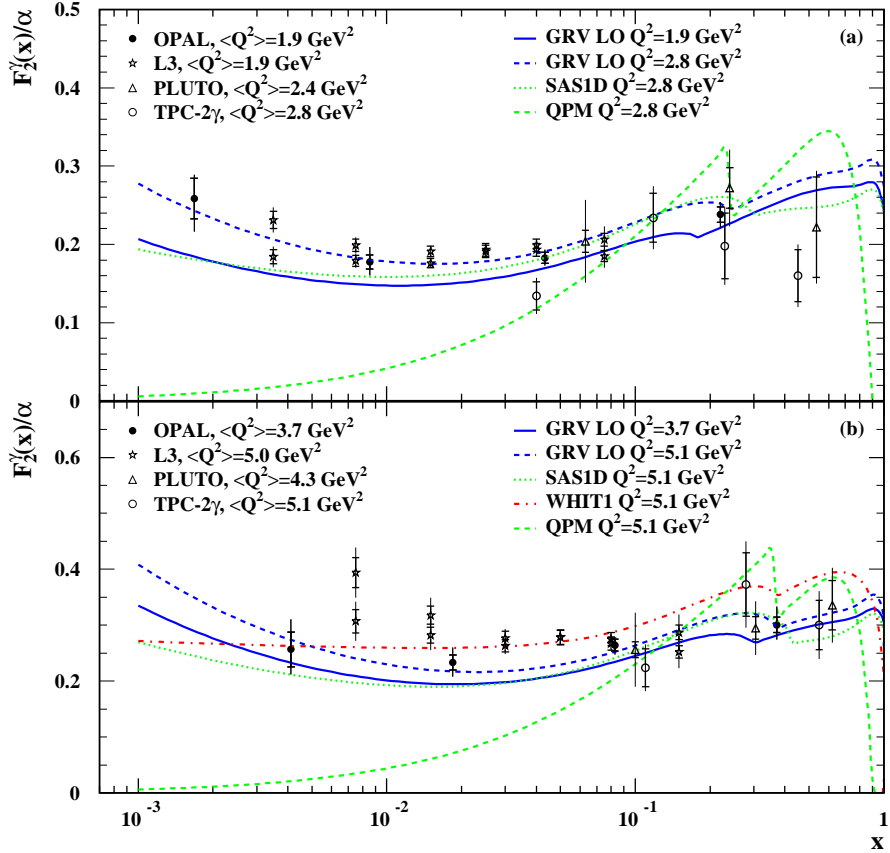}
\end{center}
\caption{ The measurement of \ftn\ 
for \qzm\ values of (a) 1.9 and (b) 3.7~\gevsq. Shown are the
results from OPAL, L3, PLUTO, and
TPC/2$\gamma$. For L3 the two sets of points were
unfolded using different Monte Carlo programs. The curves show the GRV LO,
SaS1D, WHIT1 and QPM structure functions.}
\label{fig:f2g}
\end{figure}

The photon is the gauge boson of the electromagnetic force and
therefore has fundamental direct couplings to all charged
particles. However, it can fluctuate into hadronic
matter before it takes part in an interaction, thereby giving rise
to a wide range of phenomena which can be quantified in observables.
Once measured, they can be used to improve our
understanding of both perturbative and non-perturbative QCD.
In this paper the experimental environment is provided by
photon-photon collision events recorded by the four LEP collaborations.
Scattering events of this type are commonly characterised by the
absolute four momentum squared of the photons participating in the
interaction, $Q^2$ and $P^2$, the invariant mass $W_{\gamma\gamma}$ of
the resulting hadronic final state and the Bjorken variable 
$x=Q^2/(Q^2 + W_{\gamma\gamma}^2 + P^2)$. 
For large enough virtuality $Q^2 (P^2)$ one
(both) electron(s) has a scattering angle large enough to be detected in
the experiment. Such events are referred to as single (double) tag
events. The observables studied are the structure function of the
photon \ftg\ at low $x$ and the production of heavy quarks in untagged
and single-tagged photon-photon scattering events.

\section{The Photon Structure Function \ftg}

\begin{figure}[t]
\begin{center}
\includegraphics[width=0.48\textwidth, trim= 0 30 0 0]{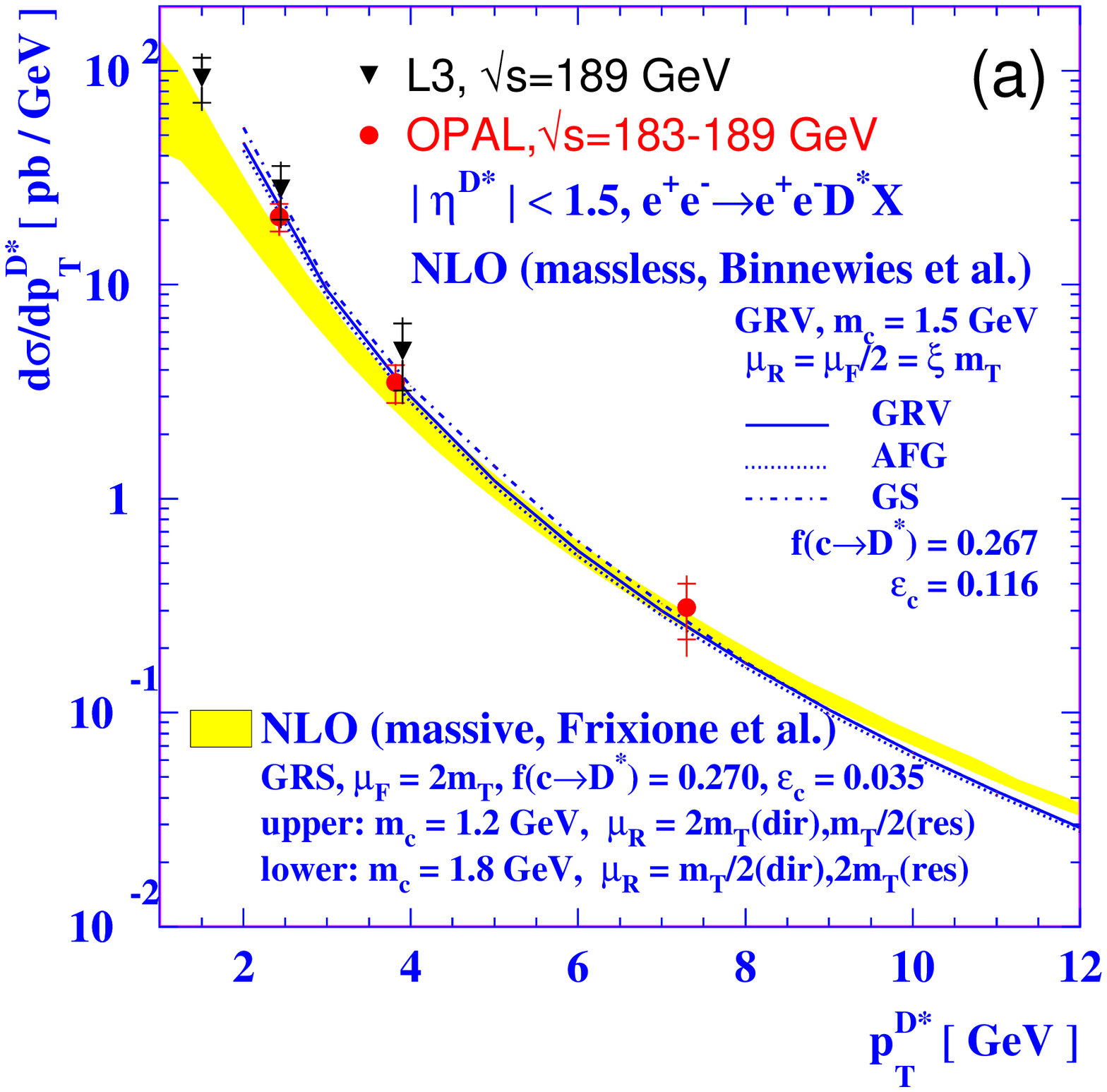}
\includegraphics[width=0.45\textwidth]{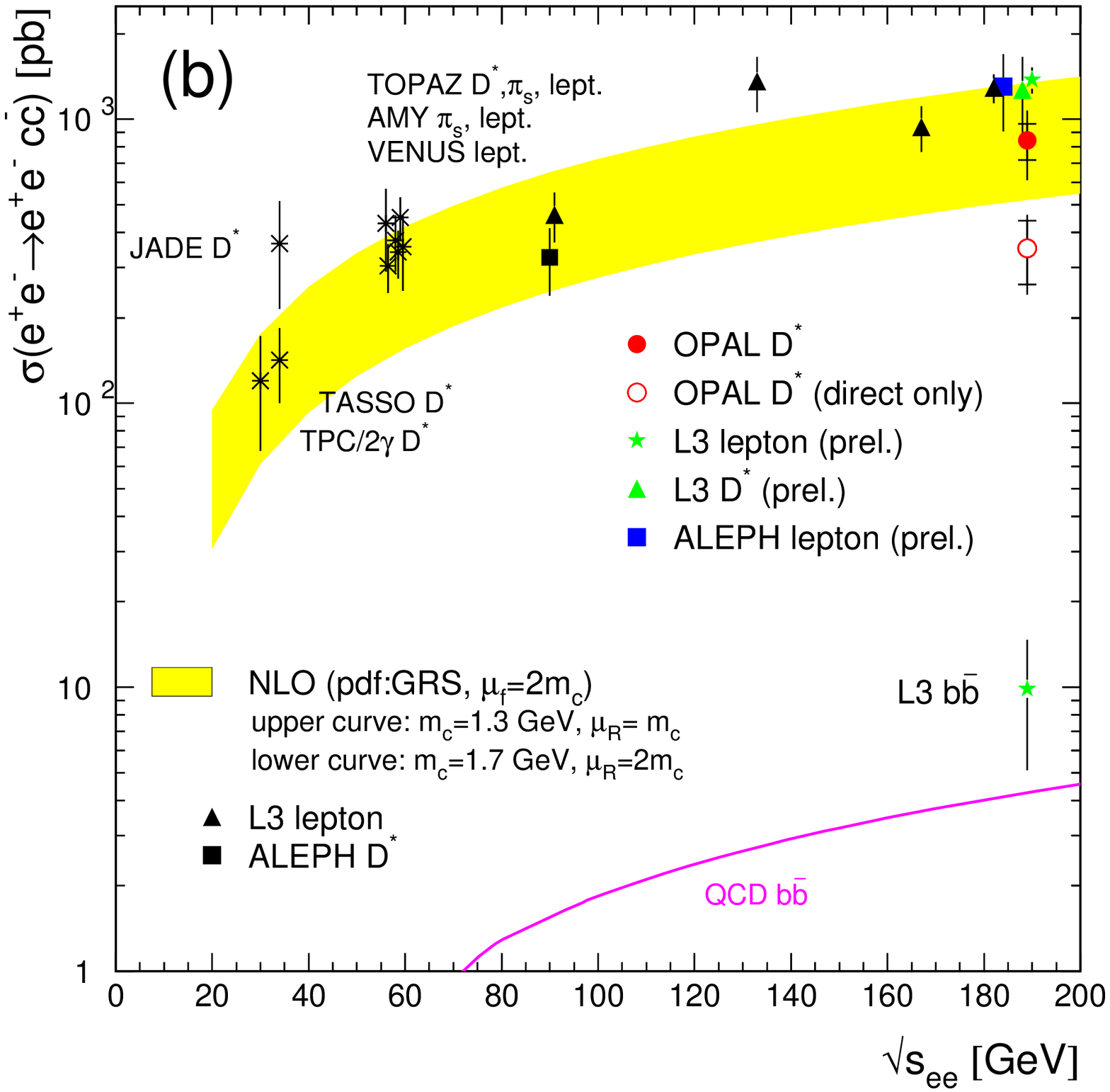}
\end{center}
\caption{(a) The differential $\DST$ cross-section, (for
         anti-tagged events) obtained by OPAL and L3. 
         The data are compared to a NLO calculation by
         Binnewies et al. 
         and to a NLO calculation by Frixione et al. using the
         massive approach. Figure (b) shows
         the measured cross-sections for the process
         \epem $\to$ \epem $\ccbar$ where the charm quarks are
         produced in the collision of two quasi-real photons. The band
         shows a NLO calculation  
         of the process \epem $\to$ \epem $\ccbar$~. 
         Also shown is the
         $\bbbar$ cross section as measured by L3.} 
\label{fig:dstar}
\end{figure}

The differential cross section of deep-inelastic electron-photon
scattering, as can be obtained from single tag events, can be
expressed as
\begin{equation}
     \frac{d^2\sigma_{{\rm e}\gamma\rightarrow{\rm e}{\rm X}}}{dxdQ^2}
   = \frac{2\pi\aemsq}{x\,Q^{4}}
     \left[\left( 1+(1-y)^2\right) \ftxq - y^{2} \flxq\right]\, ,
\end{equation}
where $\alpha$ is the fine structure constant and $y$ is the
inelasticity. In the region of small $y$ under study the contribution
from \flxq\ can be neglected and one obtains the photon structure
function \ftg. The latest measurements in the low-$x$ region come from
LEP, namely L3 and OPAL, as shown in Figure~\ref{fig:f2g}~\cite{bib-f2g}. 
For mean
values of \qzm\ = 1.9~\gevsq\ values of $x$ down to $2\cdot10^{-3}$ have
been reached. Particular emphasis is put on the behaviour of \ftg\ at
low $x$, since, driven by the hadron-like properties of the
interacting photon, one suspects a rise of the photon structure
function towards small $x$, similar to the rise observed for the
proton structure function \ftp.

\noindent
The precision of the measurement at low $x$ has been
considerably improved and there appears to be mounting evidence for
the expected rise of \ft\ at low $x$ although a final conclusion
should not be drawn yet.
The GRV LO and SaS1D parameterisations are generally consistent with the
data in all the accessible $x$ and \qsq\ regions, with the exception of
the measurement at the lowest scale, $\qzm=1.9$~\gevsq, where GRV is too low.
In contrast, the naive quark-parton model is not able to describe the
data for $x<0.1$. These results show that the photon must contain a 
significant hadron-like component at low $x$.

\begin{figure}[t]
\begin{center}
\includegraphics[width=0.95\textwidth]{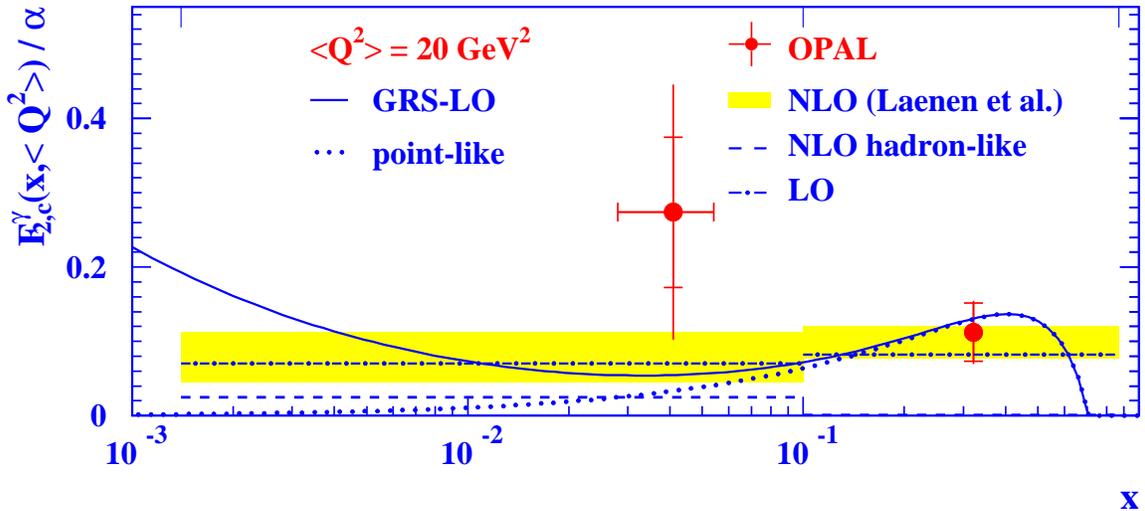}
\end{center}
\caption{The charm structure function of the photon divided
         by the fine structure constant, $\XQCF2/\alpha$, for
         an average $\langle Q^2\rangle$ of $20~{\rm GeV}^2$.
         The data are compared to the calculation of
         Laenen et al. performed in LO and NLO.
         Also shown are the prediction of the GRS-LO parametrisation 
         for the whole structure function at 
         $\langle Q^2\rangle =20~{\rm GeV}^2$ 
          and the point-like component alone.}

\label{fig:f2c}
\end{figure}

\section{Heavy Quark Production}

Charged $\DST$ mesons
provide a clean tag to study open charm production in
photon-photon collisions.
The inclusive cross-section for the production
of $\DST$ mesons can be calculated in NLO
perturbative QCD (pQCD) and this process is therefore suitable to test
the validity of the theory in this region. L3 and OPAL have determined 
the differential cross-sections ${\rm d}\sigma/{\rm d}\ptdst$
in anti-tagged \epem$\to$ \epem $\DST X$ events 
as a function of the transverse momentum $\ptdst$ as shown in Figure
\ref{fig:dstar}a~\cite{bib-opdstar,bib-l3dstar}. 
Both measurements can been seen to be in agreement
with each other within the experimental uncertainties. They are 
compared to a NLO calculation by Binnewies et al.~\cite{bib-Kniehl}
using the massless approach, and by Frixione et
al.~\cite{bib-Frixione} using the massive  approach. 
It is found that despite the low values of $\ptdst$ studied 
the massless calculation is in good agreement with the data.
The massive calculation agrees with the measured cross-section for
$\ptdst>3~{\rm GeV}$ but underestimates the data for lower values of
$\ptdst$.

The total cross-section of the process \epem $\to$ \epem $\ccbar$, where
the charm quarks are produced in the collision of two quasi-real
photons, can be deduced from measurements like the above by extrapolating
to the full phase space using MC model predictions. The results for
various experiments are shown in Figure 
\ref{fig:dstar}b~\cite{bib-opdstar,bib-aldstar}. All results
are in agreement with each other where comparable, and are well
described by the NLO calculation of Ref.~\cite{bib-DKZZ} over the
whole range of \epem centre-of-mass energies covered. Also shown in
this figure is the total $\bbbar$ cross-section in photon-photon
collisions as reported for the first time by L3~\cite{bib-l3bb} to be
$\sigma^{ee\to eebbX} = 9.9 \pm 2.9({\rm stat}) \pm 3.8 ({\rm syst})$
pb. 
This first sign of $b$ production in this process has been
obtained by exploiting the more energetic leptons from $b$
semi-leptonic decays as opposed to those from charm semi-leptonic
decays. The measured value is somewhat above the QCD prediction,
although the still sizable experimental uncertainties inhibit a clear
conclusion here.

The first measurement of the charm structure function $\XQCF2$ of the photon
has been performed by OPAL~\cite{bib-opdstar} based on about 30~$\DST$ mesons 
reconstructed in single-tagged events. The value of $\XQCF2$ is determined
for an average $\langle Q^2\rangle$ of $20~{\rm GeV}^2$ and
in two regions of $x$, $0.0014<x<0.1$ and $0.1<x<0.87$.
For $x>0.1$, the perturbative NLO calculation of Laenen et
al.~\cite{bib-Laenen} is in good agreement with the measurement.
For $x<0.1$, the measurement suffers from large uncertainties of
the invisible cross-section predicted by the HERWIG and Vermaseren
Monte Carlo models, and therefore the result is not very precise.
However, the data clearly suggest a non-zero hadron-like component of $\CF2$.

\section{Conclusion}

Data is available now which explores the structure function \ftg\ of the
photon in deep-inelastic electron-photon scattering  down to values of
$x\approx2\cdot10^{-3}$. Thanks to the improved precision of the
measurements there appears to be mounting evidence for a rise of \ftg\
at low $x$, but no final conclusion can be drawn yet.
Open charm production is studied via
the measurement of charged $\DST$-mesons and found to be in good 
agreement with NLO perturbative QCD calculations for both differential
cross-sections and the total charm cross-section measured. A first
sign of $b$ production in photon-photon collisions has been reported
and is seen to be somewhat higher than the QCD prediction, although
experimental uncertainties are still large. A first measurement of the
charm structure function has been performed in single tagged events
and has been found to be in good agreement with NLO pQCD for $x>0.1$.
A hadron-like component of $\CF2$ at low $x$ is clearly favoured by
the data.

\section*{Acknowledgements}
I would like to thank the CERN SL Division for their successful
operation of the accelerator and the four LEP collaboration for
providing me with their results. Many thanks also to
S.~S\"oldner-Rembold and R.~Nisius for providing me with collective
figures on the available data and phenomenology.

\end{document}